\documentclass[conference,10pt,draftclsnofoot,onecolumn]{IEEEtran}

\usepackage{benstyle_IEEE}

\usepackage{algorithmic}
\usepackage{array}
\usepackage{url}

\newcommand{\Romnum}[1]{\uppercase\expandafter{\romannumeral#1}}

\begin{document}

%\bstctlcite{IEEEexample:BSTcontrol}

\title{Sparsity and Parallel Acquisition: Optimal Uniform and Nonuniform Recovery Guarantees}

\author{\IEEEauthorblockN{Il Yong Chun}
\IEEEauthorblockA{Department of Mathematics \\ Purdue University, USA
\\
Email: chuni@purdue.edu}
\and
\IEEEauthorblockN{Chen Li}
\IEEEauthorblockA{
Department of Mathematics \\ University of Science and Technology of China \\
Email: lichen11@mail.ustc.edu.cn
}
\and
\IEEEauthorblockN{Ben Adcock}
\IEEEauthorblockA{Department of Mathematics \\ Simon Fraser University, Canada\\
Email: ben\_adcock@sfu.ca}}


\maketitle

\begin{abstract}
The problem of multiple sensors simultaneously acquiring measurements of a single object can be found in many applications.  In this paper, we present the optimal recovery guarantees for the recovery of compressible signals from multi-sensor measurements using compressed sensing.  In the first half of the paper, we present both uniform and nonuniform recovery guarantees for the conventional sparse signal model in a so-called distinct sensing scenario.  In the second half, using the so-called sparse and distributed signal model, we present nonuniform recovery guarantees which effectively broaden the class of sensing scenarios for which optimal recovery is possible, including to the so-called identical sampling scenario.  To verify our recovery guarantees we provide several numerical results including phase transition curves and numerically-computed bounds.
\end{abstract}

% \fi
\IEEEpeerreviewmaketitle

\section{Introduction}
In compressed sensing (CS) it is conventional to consider recovery of an $s$-sparse signal $x \in \bbC^{N}$ from single-sensor measurements of the form
\be{
\label{standard_CS}
y = A x +e,
}
where $A \in \bbC^{m \times N}$ and $e \in \bbC^{m}$ is noise.  As is well-known, for appropriate (i.e.\ incoherent) $A$, exact recovery of $x$ is possible with a number of measurements scaling linearly in $s$.  In this paper, we consider the extension of \R{standard_CS} to a multi-sensor CS problem \cite{Chun&Adcock:16arXiv-CS&PA} wherein the measurements take the form
\be{
\label{parallel_CS}
y = A x + e,\qquad A = \left [ \begin{array}{c} A_1 \\ \vdots \\ A_C \end{array} \right ],\qquad y =  \left [ \begin{array}{c} y_1 \\ \vdots \\ y_C \end{array} \right ],\qquad e =  \left [ \begin{array}{c} e_1 \\ \vdots \\ e_C \end{array} \right ].
}
Here $A_{c} \in \bbC^{m_c \times N}$ is the matrix corresponding to the measurements taken in the $c^{\rth}$ sensor and $e_c \in \bbC^{m_c}$ is noise.  Throughout this paper, we assume that the measurement matrices in the individual sensors take the form
\bes{
A_c = \tilde{A}_{c} H_c,
}  
where $\tilde{A}_{c} \in \bbC^{m_c \times N}$ are standard CS matrices (e.g.\ a random subgaussian, subsampled isometry or random convolution), and $H_c \in \bbC^{N \times N}$ are fixed, deterministic matrices, referred to as \textit{sensor profile} matrices.  These matrices model environmental conditions in the sensing problem; for example, a communication channel between $x$ and the sensors, the geometric position of the sensors relative to $x$, or the effectiveness of the sensors to $x$.  As in standard CS, our recovery algorithm will be basis pursuit:
\be{
\label{recovery_alg}
\min_{z \in \bbC^N} \| z \|_{1}\ \mbox{subject to $\| A z - y \|_2 \leq \eta$},
}
Here $\eta > 0$ is such that $\| e \|_2 \leq \eta$.   

\subsection{Applications}\label{sec:Appl}
Multi-sensor problems of the form \R{parallel_CS} arise in numerous of applications, and are used to alleviate a variety of problems in single-sensor acquisition.  In Magnetic Resonance Imaging (pMRI), for example, parallel acquisition is often employed over single-coil MRI techniques to reduce scan duration.  The most general system model in pMRI can be formulated as \R{parallel_CS} \cite{Chun&Adcock&Talavage:15TMI, Chun&Adcock&Talavage:14EMBS_pMRI, Pruessmann&Weiger&Scheidegger&Scheidegger:99MRM}, with $H_c$ being diagonal matrices corresponding to the coil sensitivity profiles.

On the other hand, in multi-view imaging $C$ cameras with differing alignments simultaneously image a single object, thus allowing a higher-resolution or higher-dimensional image to be recovered.  Applications include satellite imaging, remote sensing, three-dimensional imaging, super-resolution imaging and more.  Following the work of \cite{Park&Wakin:12EJASP, Traonmilin&etal:15JMIV}, this can be understood in terms of the above framework, with the sensor profiles $H_c$ corresponding to geometric features of the scene.

The well-known problem of derivative sampling -- with applications to super-resolution and seismic imaging -- can also be viewed in terms of \R{parallel_CS}.  The benefits of a multi-sensor system in this setting are in reducing the total cost of the acquisition problem or in enhancing the accuracy of the recovered images.  Similarly, in wireless sensor networks, a parallel acquisition setup may be used to reduce the total power consumption.  

For further discussion and applications, see \cite{Chun&Adcock:16arXiv-CS&PA}.

\subsection{Contributions}
In this paper, building on our previous work \cite{Chun&Adcock:16arXiv-CS&PA,Chun&Adcock:16ITW}, we present a series of recovery guarantees for \R{parallel_CS}--\R{recovery_alg}.  Throughout, our aim is to determine optimal measurement conditions which depend linearly on the sparsity $s$ and are independent of the number of sensors $C$.  If this holds, one confirms the benefit of a multi-sensor system over a single-sensor system, since the average number of measurements per sensor $m/C$ decreases linearly with $C$.  

In the first part of the paper, we consider a \textit{distinct} sampling scenario.  In this setting, the matrices $\tilde{A}_1,\ldots,\tilde{A}_C$ are independent; that is, drawn independently from possibly different distributions.  We present both a nonuniform recovery guarantee and a new uniform recovery guarantee for the sparse signal model.  In the second part of the paper we address the more challenging scenario of \textit{identical} sampling, wherein $m_1 = \ldots = m_C = m/C$ and the matrices $\tilde{A}_{1} = \ldots = \tilde{A}_C = \tilde{A} \in \bbC^{m/C \times N}$.  In other words, the measurement process in each sensor is identical, the only difference being in the sensor profiles $H_c$.  Using the so-called sparse and distributed signal model, we present a nonuniform recovery guarantee for this problem.  Finally, we confirm our theoretical results via phase transition curves.

\subsection{Notation}
We write $\nm{\cdot}_{p}$ for the vector $p$-norm and $\| \cdot \|_{p \rightarrow p}$ for the matrix $p$-norm.  If $\Delta \subseteq \{1,\ldots,N\}$ then we write $P_{\Delta}$ for the orthogonal projection $P_{\Delta} \in \bbC^{N \times N}$ with $(P_{\Delta}x)_{j} = x_j$, $j \in \Delta$, and $(P_{\Delta} x)_j = 0$ otherwise.  We also use the notation $A \lesssim B$ or $A \gtrsim B$ to mean there exists a constant $c>0$ independent of all relevant parameters (in particular, the number of sensors $C$) such that $A \leq c B$ or $A \geq c B$ respectively.  

A vector $z \in \bbC^N$ is $s$-sparse for some $1 \leq s \leq N$ if $\| z \|_0 = | \{ j : z_j \neq 0 \} | \leq s$.  We write $\Sigma_{s}$ for the set of $s$-sparse vectors and, for an arbitrary $x \in \bbC^N$, write $\sigma_{s}(x)_1 = \min \left \{ \| x - z \|_{1} : z \in \Sigma_{s} \right \}$, for the error of the best $\ell_1$-norm approximation of $x$ by an $s$-sparse vector.

\section{Distinct sampling and sparse vectors: Nonuniform recovery} \label{sec:nunif_dist_sparse}

\subsection{Setup}
\label{sec:setupDist2}
Our setup in this section is based on ideas from \cite{Candes&Plan:11IT} for single-sensor CS.  Suppose that $G$ is a distribution of vectors in $\bbC^N$.  We say that $G$ is isotropic if
\bes{
\bbE(a a^*) = I,\qquad a \sim G,
}
and we define the coherence $\mu(G)$ to be the smallest constant such that $\| a \|^2 \leq \mu(G)$ almost surely for $a \sim G$.

Suppose now that $G_1,\ldots,G_C$ are isotropic distributions of vectors in $\bbC^N$, where $G_c$ represents the sensing in the $c^{\rth}$ sensor, and define $F_1,\ldots,F_C$ so that $a_c \sim F_c$ if $a_c = H^*_c \tilde{a}_c$ for $\tilde{a}_c \sim G_c$.  We assume that the matrices $H_c$ satisfy the \textit{joint isometry condition}
\be{
\label{joint_iso_distinct}
C^{-1} \sum^{C}_{c=1} H^*_c H_c = I.
}
For each $c$, draw $a_{c,1},\ldots,a_{c,m/C}$ i.i.d.\ from $F_c$ and form the measurement matrix
\be{
\label{A_dist_def}
A = \frac{1}{\sqrt{m}} \left [ \begin{array}{c} A_1 \\ \vdots \\ A_C \end{array} \right ],\qquad A_c = \left [ \begin{array}{c} a^{*}_{c,1} \\ \vdots \\ a^{*}_{c,m/C} \end{array} \right ],\qquad \ c=1,\ldots,C.
}
Note that this setup allows us to consider a wide range of different sensing vectors $\tilde{a}_c$, including not only subgaussian random sensing, but also subsampled isometries (e.g.\ subsampled DFT) and random convolutions \cite{Candes&Plan:11IT}.

\subsection{Nonuniform recovery guarantee}
Our first recovery guarantee is the following:

\thm{[Nonuniform recovery for distinct sampling with sparsity model]
\label{c:distinct_sparsity}
Let $x \in \bbC^{N}$, $0 < \epsilon < 1$ and $N \geq s \geq 2$.  Suppose that $H_1,\ldots,H_C$ satisfy \R{joint_iso_distinct} and draw $A$ according to \S \ref{sec:setupDist2}.  If $y = A x + e$ with $\| e \|_2 \leq \eta$, then for any minimizer $\hat{x}$ of
\bes{
\min_{z \in \bbC^N} \| z \|_{1}\ \mbox{subject to $\| A z - y \|_2 \leq \eta$},
}
we have
\bes{
\| x - \hat{x} \|_{2} \lesssim \sigma_{s}(x)_1 + \sqrt{s} \eta,
}
with probability at least $1-\epsilon$, provided
\bes{
\label{eq:meas_nunif_recov_concent}
m \gtrsim s \cdot \mu \cdot \left ( \max_{c=1,\ldots,C} \| H_{c} \|^2_{1 \rightarrow 1} \right ) \cdot L,
}
where $\mu =  \max_{c=1,\ldots,C} \mu(G_c)$.
}
\prf{
Corollary 3.1 of \cite{Chun&Adcock:16arXiv-CS&PA} gives that the conditions of the result hold, provided $m \gtrsim s \cdot \left ( \max_{c=1,\ldots,C} \mu(F_c) \right ) \cdot L$.  Let $a_{c} \sim F_c$ and write $a_c = H^*_c \tilde{a}_c$ where $\tilde{a}_c \sim G_c$.  Then
\bes{
| (a_{c})_i | \leq \sum^{N}_{j=1} | (H_c)_{j,i} | | (\tilde{a}_{c})_j | \leq \| \tilde{a}_c \|_{\infty} \| H_c \|_{1 \rightarrow 1}.
}
Thus $\mu(F_c) \leq \mu(G_c) \| H_c \|^2_{1 \rightarrow 1}$, as required.
}

This result is \textit{nonuniform} in the sense that each random draw of $A$ guarantees recovery of a fixed vector, as opposed to all vectors, which is the case in a uniform guarantee.

\subsection{Examples} \label{sec:egs_nunif}
Theorem \ref{c:distinct_sparsity} asserts recovery from a number of measurements that is independent of $C$, provided $\| H_{c} \|_{1\rightarrow1} \lesssim1$.  In other words, when this condition holds, the number of measurement per sensor (equal to $m/C$) scales like $1/C$.  To elaborate, we now consider several examples of different sensor profiles $H_c$.  As discussed in \cite[\S \Romnum{1}-B]{Chun&Adcock:16arXiv-CS&PA}, the environmental conditions encompassed by the $H_c$ can often be modelled by either diagonal or circulant structures.  Hence these will form the primary examples in this paper.

\subsubsection{Diagonal sensor profiles}
Suppose that $H_c = \mathrm{diag}(h_c) \in \bbC^{N \times N}$ and let $\| h_c \|_{\infty} = \max_{i=1,\ldots,N} | h_{c,i} |$.  Then $\| H_c \|_{1 \rightarrow 1} = \| h_c \|_{\infty}$.  Hence if
\bes{
\max_{c=1,\ldots,C} \| h_c \|_{\infty} \lesssim 1,
}
we obtain an optimal recovery guarantee.  Note that the sensor profiles must satisfy \R{joint_iso_distinct}, i.e.\ $\sum^{C}_{c=1} | h_{c,i} |^2 = C$, $\forall i$.  In particular, $1 \leq \| h_c \|^2_{\infty} \leq C$, $\forall c$.

\subsubsection{Circulant sensor profiles}
Let $H_c \in \bbC^{N \times N}$ be circulant matrices with symbols $h_{c} \in \bbC^{N}$.  Then $\| H_c \|_{1\rightarrow1} = \| h_c \|_1$.  Hence, if 
\bes{
\max_{c=1,\ldots,C} \| h_c \|_{1} \lesssim 1,
}
we achieve an optimal recovery guarantee.  Note that if all the entries of $h_c$ have the same sign, then the joint isometry condition \R{joint_iso_distinct} implies that $1 \leq \| h_c \|^2_{1} \leq C$, $\forall c$.

\section{Distinct sampling and sparse vectors: uniform recovery for subgaussian sensing matrices} \label{sec:unif_dist_sparse}
We now specialize the setup of \S \ref{sec:setupDist2} to the case of subgaussian sensing vectors $\tilde{a}_c$, so that the matrices $A_c$ are of the form $A_c = \tilde{A}_c H_c$ where $\tilde{A}_c \in \bbR^{m/C \times N}$ are subgaussian random matrices (possibly with different subgaussian parameters).  Our aim is to prove a uniform recovery guarantee based on a concentration inequality for the matrix $A$ (Lemma \ref{l:conc_ineq}).

\subsection{Uniform recovery guarantee}
We first recall the following standard definition (see, for example, \cite[Def. 9.4]{Foucart&Rauhut:book}):

\defn{[Isotropic subgaussian random vector]
A random vector $Y$ on $\bbR^N$ is isotropic if $\bbE | \ip{Y}{x} |^2 = \| x \|^2$ for all $x \in \bbR^N$.  Furthermore, if for all $x \in \bbR^N$ with $\| x \|=1$ the random variable $\ip{Y}{x}$ is subgaussian with subgaussian parameter $\alpha > 0$ independent of $x$
\footnote{
A random variable is subgaussian if $\bbP(|X| \geq t ) \leq \beta \E^{-\kappa t^2}$ for $t>0$, and subexponential if $\bbP(|X| \geq t ) \leq \beta \E^{-\kappa t}$ for $t>0$.
Recall also that a mean zero random variable $X$ is subgaussian if and only if $\bbE(\exp(\theta X)) \leq \exp(\mu \theta^2)$, $\forall \theta \in \bbR$.  In this case, one has $\beta = 2$ and $\kappa = 1/(4 \mu)$.
}
, i.e.
\be{
\label{subgauss_def}
\bbE(\exp(\theta \ip{Y}{x}) ) \leq \exp(\alpha \theta^2),\qquad \forall \| x \|=1,\ \forall \theta \in \bbR,
}
then $Y$ is referred to as a subgaussian random vector.
}

\lem{[Concentration inequality for subgaussian sensing]
\label{l:conc_ineq}
For each $c=1,\ldots,C$, let $\tilde{A}_c \in \bbR^{m/C \times N}$ be a random matrix with independent, isotropic, and subgaussian rows with the same subgaussian parameter $\alpha_c$ in \R{subgauss_def}.  Let $H_c \in \bbR^{N \times N}$ satisfy the joint isometry condition \R{joint_iso_distinct} and suppose that $A$ is as in \R{A_dist_def}.
Then, for all $x \in \bbR^N$ and $0 < t <1$, we have
\bes{
\label{eq:concent_Ineq}
\bbP \left ( \left | \nmu{ A x }^2 - \| x \|^2 \right |\geq t \| x \|^2 \right ) \leq 2 \exp \left( - \zeta t^2 m \right )
}
where 
\bes{
\zeta = \left ( 32 \alpha^2_{\max} \Xi^2_{\mathrm{dist}} \max \left \{ 2 , \exp(1/(4\alpha_{\min})) \right \} + 8 \alpha_{\max} \Xi_{\mathrm{dist}} \right )^{-1}.
}
$\alpha_{\max} = \max_{c=1,\ldots,C} \{ \alpha_c \}$, $\alpha_{\min} = \min_{c=1,\ldots,C} \{ \alpha_c \}$ and $\Xi_{\mathrm{dist}} = \max_{c=1,\ldots,C} \| H_c \|^2_{2 \rightarrow 2}$.
}

\prf{
Suppose that $\|x \|=1$ without loss of generality.  Let $\tilde{a}_{c,i} \in \bbR^{N}$, $i=1,\ldots,m/C$, denote the rows of $\tilde{A}_c$ and define
\bes{
Z_{c,i} = | \ip{\tilde{a}_{c,i}}{H_c x} |^2 - \| H_c x \|^2,
}
for $i=1,\ldots,m/C$ and $c=1,\ldots,C$.
Since $\tilde{a}_{i,c}$ is isotropic,
\bes{
\bbE(Z_{c,i}) = \| H_c x \|^2 - \| H_c x \|^2 = 0.
}
Also, since $C^{-1}\sum^{C}_{c=1} H^*_c H_c = I$, we have
\eas{
\nmu{ A x }^2 - \| x \|^2 
&= m^{-1} \sum^{C}_{c=1} \sum^{m/C}_{i=1} |\ip{\tilde{a}_{c,i}}{H_c x} |^2 - C^{-1}\sum^{C}_{c=1} \| H_c x \|^2 
\\
&= m^{-1} \sum^{C}_{c=1} \sum^{m/C}_{i=1} Z_{c,i}.
}
Hence
\bes{
\bbP \left ( \left | \nmu{ A x }^2 - \| x \|^2 \right |\geq t \right ) = \bbP \left ( \left | \sum^{C}_{c=1} \sum^{m/C}_{i=1} Z_{c,i} \right | \geq m t  \right ).
}
We first note that the $Z_{c,i}$'s are independent, due to independence of the $\tilde{a}_{c,i}$'s. 
We now claim that the $Z_{c,i}$'s are subexponential random variables.  To see this, we first show that $\ip{\tilde{a}_{c,i}}{H_c x}$ is a subgaussian random variable.
If $H_c x = 0$ for some $c$, then $Z_{c,i} = 0$ for $i=1,\ldots,m/C$.  Otherwise, if $H_c x \neq 0$, we proceed as follows. 
Note that
\bes{
\bbE\left ( \exp(\theta \ip{\tilde{a}_{c,i}}{H_c x} ) \right ) 
\leq \exp(\alpha_c \theta^2 \| H_c x \|^2 ) ,
}
since $\tilde{a}_{c,i}$ is isotropic and subgaussian.  
Thus, $\ip{\tilde{a}_{c,i}}{H_c x}$ is a subgaussian random variable with parameters $\beta= 2$ and $\kappa_c = (4 \alpha_c \| H_c x \|^2 )^{-1}$.  
We now show that $Z_{c,i}$ is subexponential with parameters
\be{
\label{Zci_parameters}
\beta_c = \max \! \left\{ 2 ,  \exp( 1/(4\alpha_c) ) \right\}, \qquad \ \kappa_c = \left( 4 \alpha_c \| H_c \|^2_{2 \rightarrow 2} \right)^{-1}.
}
To see this, observe that
\eas{
\bbP ( |Z_{c,i} | \geq t ) &= \bbP \left ( \left | | \ip{\tilde{a}_{c,i}}{H_c x} |^2 - \| H_c x \|^2 \right | \geq t \right )
\\
& = \bbP \left ( | \ip{\tilde{a}_{c,i}}{H_c x} |^2 \geq t + \| H_c x \|^2 ~\cup \right.
\\
& \qquad~ \left. | \ip{\tilde{a}_{c,i}}{H_c x} |^2 \geq \| H_c x \|^2 - t \right ).
}
If $t > \| H_c x \|^2$ then
\eas{
\bbP ( |Z_{c,i} | \geq t ) 
&=  \bbP \left ( | \ip{\tilde{a}_{c,i}}{H_c x} |^2 \geq t + \| H_c x \|^2 \right ) 
\\
& \leq \bbP \left ( | \ip{\tilde{a}_{c,i}}{H_c x} | \geq \sqrt{t} \right ),
}
and, therefore, we have
\bes{
\bbP(|Z_{c,i}| \geq t ) 
\leq 2 \exp(-\kappa_c t),
}
since $\ip{\tilde{a}_{c,i}}{H_c x}$ is subgaussian.
For $0 < t \leq \| H_c x \|^2$, we have the following trivial bound:
\eas{
\bbP(|Z_{c,i} | \geq t) 
& \leq 1 \leq \exp \! \left ( \kappa_c ( \| H_c  \|^2_{2 \rightarrow 2} - t ) \right ).
}
Combining with the previous estimate, we deduce that $Z_{c,i}$ is subexponential with parameters as in \R{Zci_parameters}.

Notice that $\kappa_{c} \geq \kappa = \left ( 4 \alpha_{\max} \Xi_{\mathrm{dist}} \right )^{-1}$ and $\beta_{c} \leq \beta = \max \{ 2 , \exp(1/(4 \alpha_{\min})) \}$.  According to the Bernstein inequality for subexponential random variables \cite[Cor. 7.32]{Foucart&Rauhut:book}, it now follows that
\eas{
\bbP \left ( \left | \sum^{C}_{c=1} \sum^{m/C}_{i=1} Z_{c,i} \right | \geq m t \right ) & \leq 2 \exp \left(-\frac{ (\kappa m t)^2 / 2}{2 \beta m + \kappa m t} \right)
\\
&\leq 2 \exp \left( - \zeta mt^2 \right),
}
where in the second step we use the fact that $0 < t < 1$.
}

Recall that a matrix $A \in \bbC^{m \times N}$ satisfies the Restricted Isometry Property (RIP) of order $s$ if there exists $0 < \delta <1$ such that $(1-\delta) \| x \|^2 \leq \| A x \|^2 \leq (1+\delta) \| x \|^2$, $\forall x \in \Sigma_s$.
If $\delta = \delta_s  \in (0,1)$ is the smallest constant respectively such that RIP holds, then we refer to $\delta_s$ as the $s^{\textmd{th}}$ Restricted Isometry Constant (RIC) of $A$.  We now have the following:

\thm{[RIP based on concentration inequality]
\label{t:unif_recov_concent}
Let $A$ be as in Lemma \ref{l:conc_ineq} and $0 < \delta,\epsilon <1$.  If
\be{
\label{eq:meas_unif_recov_concent}
m \gtrsim \delta^{-2} \cdot \left ( \max_{c=1,\ldots,C} \| H_c \|^2_{2 \rightarrow 2} \right )^2 \cdot \left ( s \cdot \log(2N/s) + \log(2\epsilon^{-1}) \right ),
}
then with probability at least $1-\epsilon$, the RIC $\delta_s$ of $A$ satisfies $\delta_s < \delta$.\footnote{Note that the constant in \R{eq:meas_unif_recov_concent} implied by the symbol $\gtrsim$ depends on $\alpha_{\min}$ and $\alpha_{\max}$ (we suppress this dependence for ease of presentation).}
}

\prf{
Due to \R{joint_iso_distinct}, it follows that $C \geq \Xi_{\mathrm{dist}} \geq 1$ and therefore $\zeta^{-1} \lesssim \Xi^2_{\mathrm{dist}}$.  We now use a standard result on the RIP for matrices satisfying concentration inequalities (see \cite[Thm. 9.11]{Foucart&Rauhut:book}, for example).
}

We remark that the RIP of order $2s$ implies stable and robust recovery, uniform in $x \in \bbC^N$, when solving \R{recovery_alg}.  Hence Theorem \ref{t:unif_recov_concent} provides the first uniform recovery result for the parallel acquisition model \R{parallel_CS}--\R{recovery_alg}.  This result also gives conditions for an optimal recovery guarantee.  Provided $\Xi_{\mathrm{dist}} \lesssim 1$, the total number of measurements $m$ is independent of the number of sensors $C$.  Note that $1 \leq \Xi_{\mathrm{dist}} \leq C$ in general, as was observed in the proof of Theorem \ref{t:unif_recov_concent}.

\rem{[Universality] \label{r:univ}
Suppose that the setup in Theorem \ref{t:unif_recov_concent} is given.  If $U \in \bbR^{N \times N}$ is any deterministic orthogonal matrix, then the matrix $A U$ also satisfies the same concentration inequality as $A$.  Hence \R{eq:meas_unif_recov_concent} implies stable and robust signal recovery for not only sparsity in the canonical domain, but also sparsity in any orthogonal transform domain (e.g.\ DCT or wavelet).
}

\subsection{Examples} \label{sec:egs_unif}
As in \S \ref{sec:egs_nunif}, we now consider the case of diagonal and circulant sensor profile matrices.

\subsubsection{Diagonal sensor profiles}
When $H_c = \mathrm{diag}(h_c) \in \bbR^{N \times N}$, we have $\| H_c \|_{2 \rightarrow 2} = \| h_c \|_{\infty}$.  Therefore, if 
\bes{
\label{eq:unif_cond_diagHc}
\max_{c=1,\ldots,C} \| h_c \|^2_{\infty} \lesssim 1
}
we can obtain an optimal recovery guarantee.  Observe that this is exactly the same condition as discussed in \S \ref{sec:egs_nunif} for the nonuniform recovery guarantee.

\subsubsection{Circulant sensor profiles}
Suppose that $H_c \in  \bbR^{N \times N}$ are circulant matrices with symbols $h_c \in \bbR^{N}$.   Based on the spectral decomposition, we can write $H_c$ as $H_c = \Phi^* \Lambda_c \Phi$, where $\Phi \in \bbC^{N \times N}$ is the unitary discrete Fourier transform (DFT) matrix and $\Lambda_c = \mathrm{diag}(\lambda_c)$ is the diagonal matrix of eigenvalues of $H_c$.  Since $\| H_c \|_{2 \rightarrow 2} = \| \lambda_{c} \|_{\infty}$ and $\lambda_c = \sqrt{N} \Phi h_c$,we have $\| H_c \|_{2 \rightarrow 2} = \| \lambda_{c} \|_{\infty} \leq \| h_c \|_{1}$.  Hence, if 
\bes{
\label{eq:unif_cond_circHc}
\max_{c=1,\ldots,C} \| h_c \|_{\infty} \lesssim 1,
}
we obtain an optimal recovery guarantee.  As in the previous case, we note that this is exactly the same condition as discussed in \S \ref{sec:egs_nunif} for the nonuniform recovery guarantee.

\subsection{Discussion: Nonuniform versus uniform} 
Both the nonuniform recovery results in \S \ref{sec:nunif_dist_sparse} and the uniform recovery results in \S \ref{sec:unif_dist_sparse} assert that distinct sampling in parallel acquisition can decrease the numbers of measurements  required per sensor linearly in $C$, subject to the joint isometry condition \R{joint_iso_distinct} and specific coherence conditions on the $H_c$'s.  Interestingly, the nonuniform case stipulates a bound on the matrix $1$-norms $\| H_c \|_{1\rightarrow1}$ (Theorem \ref{c:distinct_sparsity}) whereas in the uniform case one has a bound on the matrix $2$-norms $\| H_{c} \|_{2 \rightarrow 2}$ (Theorem \ref{t:unif_recov_concent}).  For both circulant and diagonal sensor profiles, however, these result in the same conditions.

This aside, there are several other important differences between the results.  The nonuniform recovery guarantee can be applied to all types of standard CS matrices (e.g., random subgaussian, subsampled isometry or random convolution), however, it does not apply to the case where the sparsity is in a transform domain.  Conversely, the uniform recovery result considers eal subgaussian measurements only, but guarantees signal recovery when the sparsity occurs in any orthogonal transform domain (see Remark \ref{r:univ}).

\section{Beyond sparsity and distinct sampling: Nonuniform recovery based on sparse and distributed vectors}
So far, we have only considered the recovery of sparse vectors in the distinct sampling scenario.  While there are many constructions of sensor profile matrices which give provably optimal recovery guarantees, unfortunately it is also straightforward to devise reasonable sensor profiles for which optimal recovery of all sparse vectors is not possible.  A particular issue is related to clustering: namely, the possibility for the nonzeros of a sparse vector to potentially accumulate in one portion of the signal.  Certain choices of sensor profiles $H_c$ can attenuate the signal $x$ that clusters, meaning that most of the sensors give no information \cite{Chun&Adcock:16arXiv-CS&PA}.  Perhaps unsurprisingly, this situation is typically more pronounced in the case of identical sampling.  To overcome this issue, we now present (nonuniform) recovery guarantees in both the distinct and identical sampling scenarios for a more constrained signal model which prohibits such clustering.  For ease of presentation, we focus on the case of diagonal sensor profiles only in this section.

\subsection{Signal model}

We now introduce the new signal model:

\defn{[Sparsity in levels]
\label{d:sparsity_lev}
Let $\cI = \{ I_1,\ldots,I_D \}$ be a partition of $\{1,\ldots,N\}$ and $\cS = (s_1,\ldots,s_D) \in \bbN^D$ where $s_d \leq | I_d|$, $d=1,\ldots,D$.  A vector $z \in \bbC^N$ is $(\cS,\cI)$-sparse in levels if $\left | \left \{ j : z_j \neq 0 \right \} \cap I_d \right | \leq s_d$ for $d=1,\ldots,D$.
}

Note that sparsity in levels was first introduced in \cite{Adcock&Hansen&Poon&Pi&Roman:13arXiv} as a way to consider the asymptotic sparsity of wavelet coefficients (see also \cite{Adcock&Hansen&Roman:14arXiv1403.6541}).

\defn{[Sparse and distributed vectors]
\label{d:sparse_distrib}
Let $\cI = \{ I_1,\ldots,I_D \}$ be a partition of $\{1,\ldots,N\}$ and $1 \leq s \leq N$.  For $1 \leq \lambda \leq D$, we say that an $s$-sparse vector $z \in \bbC^N$ is sparse and $\lambda$-distributed with respect to the levels $\cI$ if $z$ is $(\cS,\cI)$-sparse in levels for some $\cS = (s_1,\ldots,s_D)$ satisfying
\bes{
\max_{d=1,\ldots,D} \{ s_d \} \leq \lambda s / D.
}
We denote the set of such vectors as $\Sigma_{s,\lambda,\cI}$ and, for an arbitrary $x \in \bbC^N$, write $\sigma_{s,\lambda,\cI}(x)_1$ for the $\ell_1$-norm error of the best approximation of $x$ by a vector in $\Sigma_{s,\lambda,\cI}$.
}
Note that we are interested in the case that $\lambda$ is independent of $D$; that is, when the none of $s_d$'s greatly exceeds $s/D$.

\subsection{Distinct sampling with diagonal sensor profiles} \label{sec:nonunifDist}
The setup is as in \S \ref{sec:setupDist2} except we now assume that $H_c = \mathrm{diag}(h_c)$, $h_c = \{ h_{c,i} \}^{N}_{i=1} \in \bbC^N$ are diagonal sensor profiles.

\cor{[\mbox{\cite[Cor. 3.5]{Chun&Adcock:16ITW}}]
\label{c:distinct_bound}
Let $\cI = \{ I_1,\ldots,I_D\}$ be a partition of $\{1,\ldots,N\}$, $1 \leq \lambda \leq D$, $2 \leq s \leq N$, $x \in \bbC^{N}$ and $0 < \epsilon < 1$.  Suppose that $H_1,\ldots,H_C$ are diagonal matrices satisfying the joint isometry condition \R{joint_iso_distinct} and draw $A$ as in \R{A_dist_def}.  If $y = A x + e$, $\| e \|_{2} \leq \eta$, then for any minimizer $\hat{x}$ of
\bes{
\min_{z \in \bbC^N} \| z \|_{1}\ \mbox{subject to $\| A z - y \|_2 \leq \eta$},
}
we have
\bes{
\| x - \hat{x} \|_{2} \lesssim \sigma_{s,\lambda,\cI}(x)_1 + \sqrt{s} \eta,
}
with probability at least $1-\epsilon$, provided
\bes{
\label{distinct_levels_nunif_bound}
m \gtrsim \lambda \cdot s \cdot \mu \cdot \Upsilon_{\mathrm{dist}} \cdot L,
}
where $\mu = \max_{c=1,\ldots,C} \mu(G_c)$ and
\bes{
\Upsilon_{\mathrm{dist}} =  D^{-1} \max_{c=1,\ldots,C}  \sum^{D}_{d=1} \| h_c \|_{\infty} \| P_{I_d} h_c \|_{\infty}.
}
}

\subsection{Identical sampling with diagonal sensor profiles}
The setup for identical sampling differs from that of \S \ref{sec:setupDist2}.  Let $G$ be an isotropic distribution of vectors in $\bbC^N$.  Draw $\tilde{a}_1,\ldots,\tilde{a}_{m/C}$ i.i.d.\ from $G$ and form the matrix
\bes{
\tilde{A} = \left [ \begin{array}{c} \tilde{a}^*_{1} \\ \vdots \\ \tilde{a}^*_{m/C} \end{array} \right ].
}
Now let $H_c \in \bbC^{N \times N}$ be matrices satisfying the joint isometry condition
\be{
\label{joint_iso_identical}
 \sum^{C}_{c=1} H^*_c H_c = I,
}
and form the matrix
\be{
\label{A_identical}
A = \sqrt{\frac{C}{m}} \left [ \begin{array}{c} A_1 \\ \vdots \\ A_C \end{array} \right ],\qquad A_c = \tilde{A} H_c, \qquad \ c=1,\ldots,C.
}

\cor{[\mbox{\cite[Cor. 3.6]{Chun&Adcock:16ITW}}]
\label{c:identical_bound}
Let $\cI = \{ I_1,\ldots,I_D\}$ be a partition of $\{1,\ldots,N\}$, $1 \leq \lambda \leq D$, and $2 \leq s \leq N$.  Let $x \in \bbC^{N}$, $0 < \epsilon < 1$ and $H_{c} \in \bbC^{N \times N}$, $c=1,\ldots,C$, be diagonal matrices satisfying the joint isometry condition \R{joint_iso_identical} and draw $A$ according to \R{A_identical}.  If $y = A x + e$, $\| e \|_{2} \leq \eta$, then for any minimizer $\hat{x}$ of
\bes{
\min_{z \in \bbC^N} \| z \|_{1}\ \mbox{subject to $\| A z - y \|_2 \leq \eta$},
}
we have
\bes{
\| x - \hat{x} \|_{2} \lesssim \sigma_{s,\lambda,\cI}(x)_1 + \sqrt{s} \eta,
}
with probability at least $1-\epsilon$, provided
\bes{
m \gtrsim \lambda \cdot s \cdot \mu \cdot \Upsilon_{\mathrm{idt}} \cdot L,
}
where $\mu = \mu(G)$ and
\bes{
\Upsilon_{\mathrm{idt}} = \frac{C}{D} \max_{i=1,\ldots,N} \sum^{D}_{d=1} \max_{j \in I_d} \left | \sum^{C}_{c=1} \overline{h_{c,i}} h_{c,j} \right |.
}
}

\section{Examples}
We now consider several examples of explicit sensor profile matrices to illustrate our various recovery guarantees.

\subsection{Diagonal sensor profiles}

\begin{figure}[t!]
\centering
\begin{tabular}{ccc}
\includegraphics[scale=0.55, trim=0.2em 0.4em 2.6em 2.2em, clip]{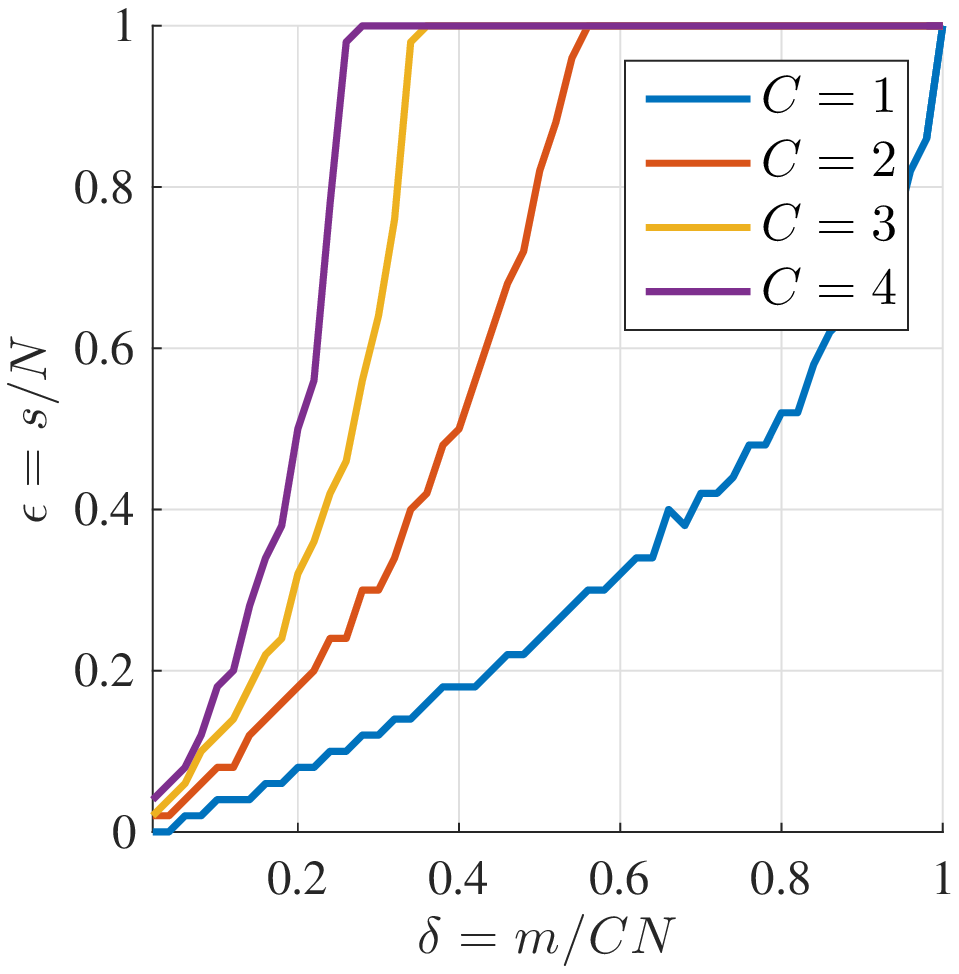} & {} &
\includegraphics[scale=0.55, trim=0.2em 0.4em 2.6em 2.2em, clip]{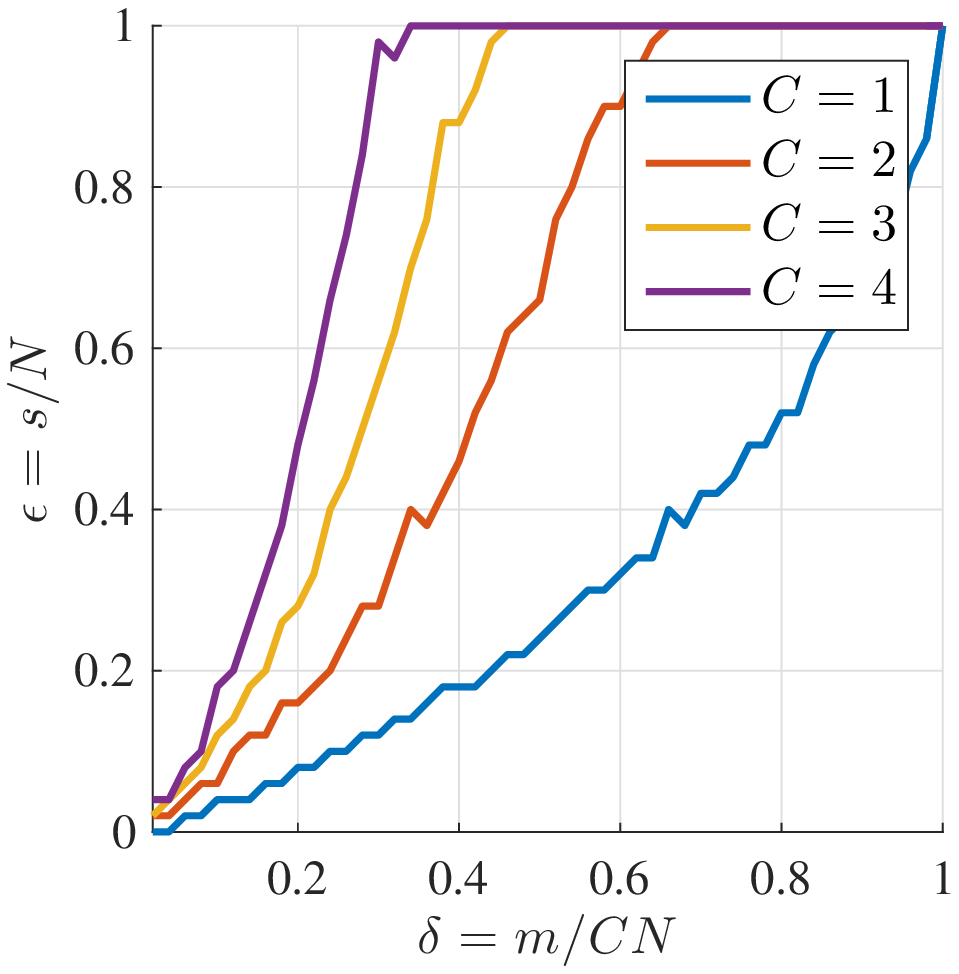} \\
{\small \qquad (a) Distinct sampling} & {\small} & {\small \qquad (b) Identical sampling} \\
\end{tabular}

\caption{
Empirical phase transitions for random Fourier sensing with banded diagonal sensor profile matrices and $C=1,2,3,4$ sensors. 
Phase transition curves with the empirical success probability $\approx 50 \%$ are presented (for details of phase transition experiment, see \cite{Chun&Adcock:16arXiv-CS&PA}).
For both sampling scenarios, the empirical probability of successful recovery increases as $C$ increases. The results are in agreement with our theoretical results. 
}
\label{fig:PTcurve_F_DiagH}
\end{figure}

\subsubsection{Piecewise constant sensor profiles}
The following example was first presented in \cite{Chun&Adcock:16arXiv-CS&PA}.  Let $\cI = \{ I_1,\ldots,I_D\}$ be a partition of $\{1,\ldots,N\}$, where $D \leq C$, and suppose that $V = \{ V_{c,d}: c=1,\ldots,C, d = 1,\ldots,D \} \in \bbC^{C \times D}$ is an isometry, i.e.\ $V^* V = I$. 
Define the sensor profile matrices $H_c = \sqrt{\frac{C}{M}} \sum^{D}_{d=1} V_{c,d} P_{I_d}$, where $M=1$ (distinct) or $M=C$ (identical), so that the $H_c$ satisfy their respective joint isometry conditions.

For distinct sampling, observe that $\| h_{c} \|^2_{\infty} \leq C \mu(V)$, where $\mu(V) = \max_{c,d} | V_{c,d} |^2$ is the coherence of $V$.  Hence we obtain optimal uniform and nonuniform guarantees (see Theorems \ref{t:unif_recov_concent} and \ref{c:distinct_sparsity} respectively) for the recovery of $s$-sparse vectors if $V$ is incoherent, i.e.\ $\mu(V) \lesssim C^{-1}$.\footnote{Note that since $V$ is an isometry, we have $C^{-1} \leq \mu(V) \leq 1$.}

Conversely, for identical sampling, if we set $D=C$ then it follows that $\Upsilon_{\mathrm{idt}} = 1$ (since $V$ is an isometry).  Hence we obtain an optimal nonuniform recovery guarantee for the sparse and distributed signal model.\footnote{It is straightforward to see that optimal recovery of all sparse vectors is not possible for identical sampling with this class of sensor profiles \cite{Chun&Adcock:16ITW}. }

\subsubsection{Banded sensor profile} 
\label{sec:eg:banded}
Let $\cI = (I_1,\ldots,I_D)$ be a partition and suppose that the $h_c$ are banded, i.e.\
\bes{
\supp(h_c) \subseteq \bigcup^{r_2}_{d=-r_1} I_{c+d},
}
for some fixed $r_1 \in \bbN$ and $r_2 \in \bbN$ (note that $I_{c+d} = 0$ if $c+d < 0$ or $c+d > C$).  
As discussed in \cite[\S \Romnum{4}-B]{Chun&Adcock:16ITW}, one obtains an optimal recovery guarantee for both distinct and identical sampling in this case with the sparse and distributed signal model with $D = C$ levels, provided $r_1+r_2$ is independent of $C$.  A specific example of this setup is a smooth sensor profile with compact support \cite[Fig.\ 1(c)]{Chun&Adcock:16arXiv-CS&PA}, which corresponds to a sharply decaying coil sensitivity in a one-dimensional (1D) example of pMRI;  see \cite{Chun&Adcock&Talavage:15TMI} for details.  The optimal recovery guarantee for this example is verified in Fig.\ \ref{fig:PTcurve_F_DiagH}(b).

\subsubsection{Global and oscillatory sensor profiles}
\begin{figure}[t!]
\centering
\begin{tabular}{c}
\includegraphics[scale=0.55, trim=0.2em 0.55em 2.6em 1.5em, clip]{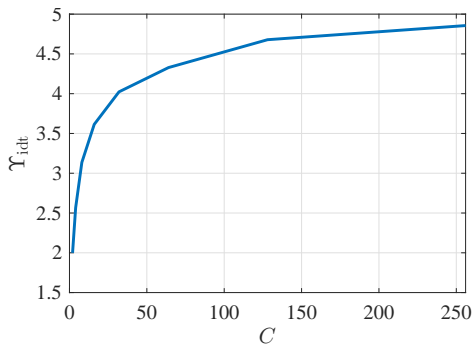} \\
\end{tabular}

\caption{
Computed $\Upsilon_{\mathrm{idt}}$ values with different numbers of $C$ ($C \in \{2,4,8,\ldots,128,256\}$, $D=C$, and $N=512$):
Different from worst case bound in \cite[Cor. 4.2]{Chun&Adcock:16arXiv-CS&PA}, $\Upsilon_{\mathrm{idt}}$ does not increase in $C$ as $C$ increases, i.e.\ the number of measurements required per sensor decreases as $C$ increases.
}
\label{fig:comptBound}
\end{figure}

As opposed to banded sensor profiles, we now consider global, oscillatory profiles of the form $h_{c,i} = \exp(2 \pi \I c i / N) / \sqrt{M}$ for $i=1,\ldots,N$ and $c=1,\ldots,C$.
These types of profiles can be used to model a wireless sensor network application in the case where the wireless channel between a source and sensors is time varying.  Since $\| h_c \|_{\infty} = 1$, we deduce optimal uniform and nonuniform sparse signal recovery guarantees for distinct sampling with these profiles.  On the other hand, for identical sampling one can compute $\Upsilon_{\mathrm{idt}}$ for different values of $C$ and $D$.  If $D = 1$ then $\Upsilon_{\mathrm{idt}}$ scales linearly with $C$, implying that optimal recovery of sparse vectors cannot be ensured.  However, as shown in Fig.\ \ref{fig:comptBound}, $\Upsilon_{\mathrm{idt}}$ remains bounded when $D = C$.  This implies an optimal recovery of sparse and distributed vectors.

\subsection{Circulant sensor profiles}
\begin{figure}[t!]
\centering
\begin{tabular}{ccc}
\includegraphics[scale=0.55, trim=0.2em 0.4em 2.6em 2.2em, clip]{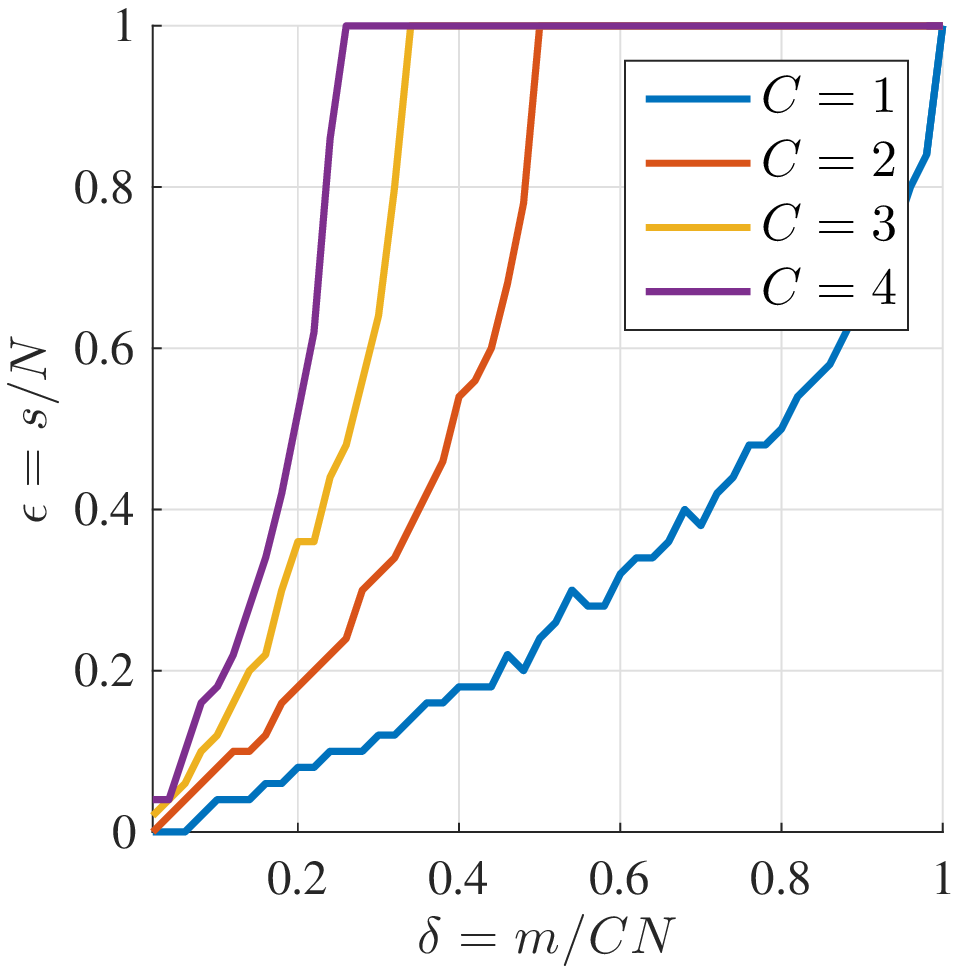} & {} &
\includegraphics[scale=0.55, trim=0.2em 0.4em 2.6em 2.2em, clip]{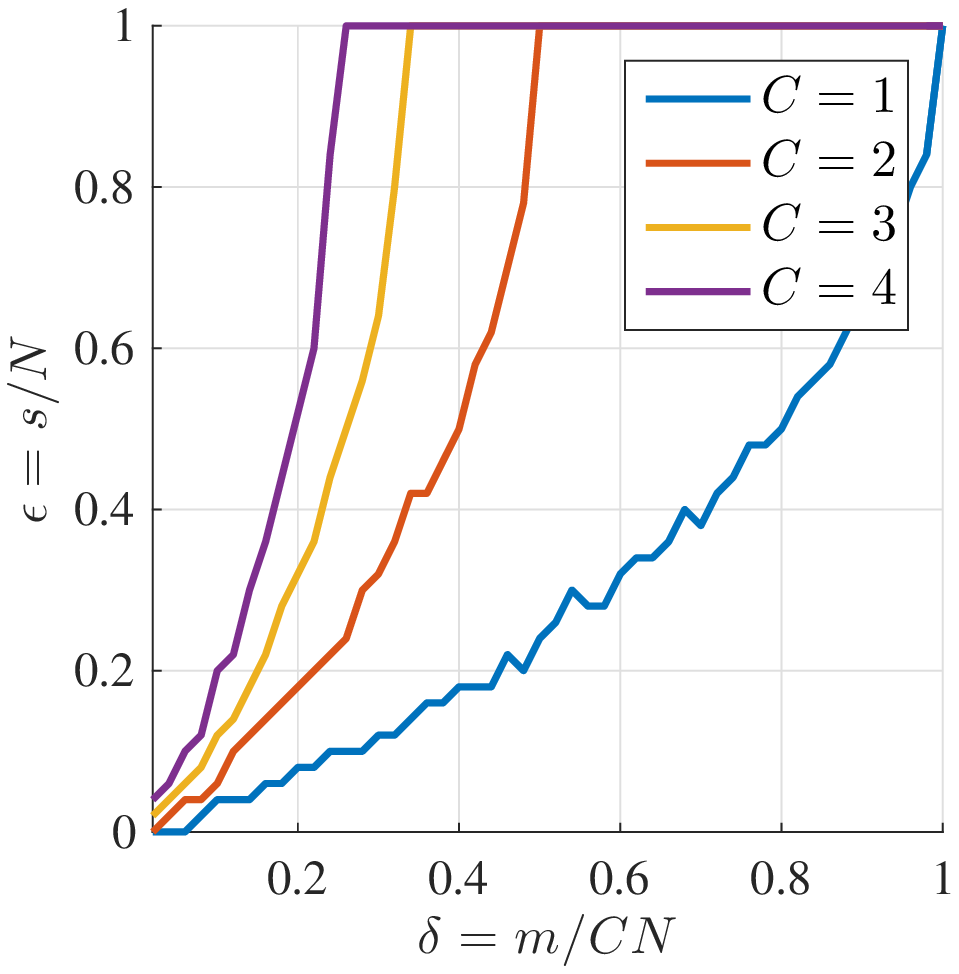} \\
{\small \qquad (a) Distinct sampling} & {\small} & {\small \qquad (b) Identical sampling} \\
\end{tabular}

\caption{
Empirical phase transitions for random Gaussian sensing with circulant sensor profile matrices and $C=1,2,3,4$ sensors. 
Phase transition curves with the empirical success probability $\approx 50 \%$ are presented (for details of phase transition experiment, see \cite{Chun&Adcock:16arXiv-CS&PA}).
For both sampling scenarios, the empirical probability of successful recovery increases as $C$ increases. The results (a) are in agreement with our theoretical results in distinct sampling. 
}
\label{fig:PTcurve_G_CircH}
\end{figure}

Finally, in Fig.\ \ref{fig:PTcurve_G_CircH}, we consider circulant sensor profile matrices, corresponding to a 1D example of the multi-view imaging application.  The circulant matrices were constructed with eigenvalues uniformly distributed on the unit circle, so that $\| h_{c} \|_{1} \lesssim 1$ where $h_c$ is the symbol of $H_c$.  As discussed in \S \ref{sec:egs_nunif} and \S \ref{sec:egs_unif}, this gives optimal uniform and nonuniform recovery guarantees in the case of distinct sampling, thus explaining the results in Fig.\ \ref{fig:PTcurve_G_CircH}(a).  Interestingly, Fig.\ \ref{fig:PTcurve_G_CircH}(b) suggests that identical sampling also exhibits an optimal recovery guarantee, although we have no proof of this fact.

\section*{Acknowledgments}
BA wishes to acknowledge the support of Alfred P. Sloan Research Foundation and the Natural Sciences and Engineering Research Council of Canada through grant 611675.  BA and IYC acknowledge the support of the National Science Foundation through DMS grant 1318894.

%%%%%%%%%%%%%%%%%%%%%%%%%%%%%%%%%%%%%%%%%%%%%%%%%%%%%%%%%%%%%%%%%%%%%%%%%%%%%%%%
%                                References
%%%%%%%%%%%%%%%%%%%%%%%%%%%%%%%%%%%%%%%%%%%%%%%%%%%%%%%%%%%%%%%%%%%%%%%%%%%%%%%%
%\bibliographystyle{IEEEbib}
\bibliographystyle{IEEEtran}
\bibliography{IEEEabrv,referenceBibs_Bobby}
%%%%%%%%%%%%%%%%%%%%%%%%%%%%%%%%%%%%%%%%%%%%%%%%%%%%%%%%%%%%%%%%%%%%%%%%%%%%%%%%
\end{document}